\documentstyle[aps,prb,preprint,graphicx,amsbsy,amssymb]{revtex}
\tightenlines

\begin{document}

\title{ Symmetry Considerations for the Detection of Second-Harmonic Generation in
     Cuprates in the Pseudogap phase.}
\author{M.E.Simon}
\author{C.M.Varma}
\address{Bell Laboratories, Lucent Technologies, Murray Hill, NJ 07974}

\maketitle
\date{\today}

\begin{abstract}

A proposal to test the proposed time-reversal and inversion breaking phase 
in the Pseudogap 
region of the Cuprate compounds
through the variation of Second-harmonic generation intensity with temperature and 
polarization and angle of incidence
is presented.

\end{abstract}

\pacs{74.20.Mn,74.25.Gz,42.65.Ky}

\section{Introduction}

     Feibig et al. \cite{feibig} have discussed the conditions for second harmonic generation proportional to the
     Time-reversal breaking order parameter for the case of {\it  magnetoelectric symmetry}\cite{LL}, i.e.
     for the case that Time-reversal Symmetry R is broken and so is Inversion I but their product
     RI is preserved. Following earlier suggestions \cite{cmv1,simon}, ARPES experiments with
     circularly polarized photons \cite{kaminski} have detected such a phase in the 
so-called {\it pseudogap phase}
     in the phase diagram of a high temperature superconducting compound of the Copper-Oxide (Cu-O)
family. Time-reversal is broken due to circulating currents
     in specific patterns within the unit cell with translational symmetry preserved.
     If this discovery is correct, it togather with the extensive evidence for {\it Quantum Critical
     properties} in the normal phase \cite{vnv} around the doping- density for the highest $T_c$ implies that
     the basic framework for the theory of the Cuprates is in place. It is therefore very important to
     verify the discovery by independent experiments. Accordingly, we work out here the symmetry
      considerations for magneto-electric induced second harmonic generation in the Cuprates.
     
    Many terms in the multipole susceptibility  contributes to the second harmonic generation (SHG). 
    The relevant contributions for the processes discussed here come from the magnetic-dipole, 
   $M_k(2\omega) \propto \chi_{ijk}^m (\omega,\omega ,2\omega) E_i(\omega )E_j(\omega )$, and the 
     electric-dipole, $E_k(2\omega) \propto \chi_{ijk}^e (\omega,\omega ,2\omega) E_i(\omega )E_j(\omega )$. 
    While the former is generally present in the symmetry of the cuprate compounds, the latter
occurs in the predicted pseudogap phase with $\chi_{ijk}^e $ transforming as an odd tensor
under time-reversal. The intereference effects due to the simultaneous presence of the
 two-time-reversal odd tensors  $\chi_{ijk}^m $ and $\chi_{ijk}^e $ are responsible for the effects
discussed here.

      \section{Symmetry of the normal Phase}

      Most $Cu-O$ compounds possess a centre of symmetry in the unit-cell. We consider the body-centered
      tetragonal crystals with symmetry group $4/mmm$($D_{4h}$),  denoting the presence of
      a four-fold axis and a mirror plane orthogonal to this axis and two other mirror planes
      or orthorhombic crystals with symmetry group $mmm$ ($D_{2h}$), (three orthogonal mirror planes only).
      Due to the centre of inversion, the third rank polar tensor $\chi_{ijk}^e$ arising from electric-
      dipole effects is zero but the third rank axial tensor $\chi_{ijk}^m$ arising from magnetic-dipole
      effects is non-zero. In other words a term in the Free-energy $\Phi$ of the form
      \begin{equation}
      \Phi = \chi_{ijk}^m E_iE_jH_k
      \end{equation}
      exists. Correspondingly a magnetization for applied electric fields
      \begin{equation}
      M_k= \chi^m_{ijk} E_iE_j
      \end{equation}
      is non-zero. 

We specify the tensors in this paper using the following co-ordinates:
      The c axis of the crystal coincides with the $\hat{z}$ direction, a and b crystalline axis (those along the nearest
neighbor Cu-Cu vector)
      are along the $(\hat{x}+\hat{y})/\surd 2$ and $(\hat{x}-\hat{y})/\surd 2$  directions.
      For the $4/mmm$ group, three independent components of $\chi_{ijk}^m $ are non-zero \cite{birss}. The vector
      {\bf M} takes the form:
      \begin{equation}
              {\bf M} =  \left( \begin{array}{c}
                  -\chi^m_1 E_y E_z  \\
                   \chi^m_1 E_z E_x  \\
                   0 \\ 
                   \end{array} \right)
      \end{equation}
      \begin{eqnarray}
      \chi^m_1 \equiv (\chi_{xzy}^m + \chi_{zxy}^m)
      \end{eqnarray}
   
   For the $mmm$ group, six independent elements of $\chi_{ijk}^m $ exist and 
all three cartesian components of
      ${\bf M}$ are bilinear in $E^\prime s$. We present detailed results for the
$4/mmm$ crystalline symmetry. Very similar conclusions hold for the $mmm$
 crystalline symmetry.       

      \section{Time-Reversal Breaking State}
       
Two different phases breaking time-reversal are shown to be possible in 
the pseudogap phase (\cite{cmv1}, \cite{simon}). The current patterns 
in these phases in the Cu-O planes are
shown in Fig. (1).
 
One called the
       $\Theta_{II}$ phase has the symmetry $\underbar{m}mm$ (Fig. 1 b). This group is in the {\it Magnetoelectric}
      class, where both time-reversal ($R$) and inversion ($I$) are broken but their product is preserved. 
      The other possible phase  $\Theta_I$ (Fig. 1 a) preserves inversion while breaking
      Time-reversal. It has the symmetry $4/mm\underbar{m}$. So the effects discussed
       here do not occur in such a phase.

As will be clear from Fig. (1 b), the point group ${\underbar m}mm$ has the following elements:
the identity,  two-fold axis rotation around  $\hat{y}$ ($C_y$), the reflections 
in the $x-y$ and $y-z$ planes ($\sigma_z$ and $\sigma_x$), $RI$, $R\sigma_y$, $R C_x$ 
and $R C_z$.

       In the group $\underbar{m}mm$, axial tensors of third rank which are invariant under time-reversal
       $\chi_{ijk}^{m,(i)}$
       are allowed \cite{birss}.
       The independent components of ${\bf M} $ are
 
        \begin{equation}
       {\bf M}= \left( \begin{array}{c}
                    \chi^m_{x}E_yE_z \\
                   \chi_{y}^m E_xE_z \\
                   \chi_{z}^m E_xE_y \\
                   \end{array} \right).
        \end{equation}
       where $\chi_{jki}^{m} + \chi^m_{kji}= \chi^m_{i}$ and $i,j,k$ run over $x,y,z$. 
      But more importantly, polar-tensors of odd-rank $\chi_{ijk}^{e(c)}$,
        which change sign under time-reversal are also allowed.  So a Polarization vector
        of the form 
    \begin{equation}
        {\bf P} = \left( \begin{array}{c}
                    \chi_{1}^{e(c)}E_xE_y \\
                    \chi_{zzy}^{e(c)}E_z^2 + \chi_{xxy}^{e(c)}E_x^2 +
                    \chi_{yyy}^{e(c)}E_y^2\\
                    \chi_{2}^{e(c)}E_zE_y\\
                    \end{array} \right)
        \end{equation}
       is allowed.  Here  $\chi_{1}^{e(c)}=\chi_{yxx}^{e(c)} + \chi_{xyx}^{e(c)}$ and
             $\chi_{2}^{e(c)}=\chi_{yzz}^{e(c)} + \chi_{zyz}^{e(c)}$. These $\chi's$ are 
         expected to be proportional to the order parameter $\Theta_{II}$. 
         
    From these we can construct the relevant part of the Poynting vector ${\bf S}$:
          \begin{eqnarray}
          {\bf S} = \frac{\partial^2 {\bf P}}{\partial t^2} + \nabla \times \frac{\partial {\bf M}}
          {\partial t}
          \label{s}
          \end{eqnarray}

\begin{figure}[ttt]
\centering
\includegraphics[]{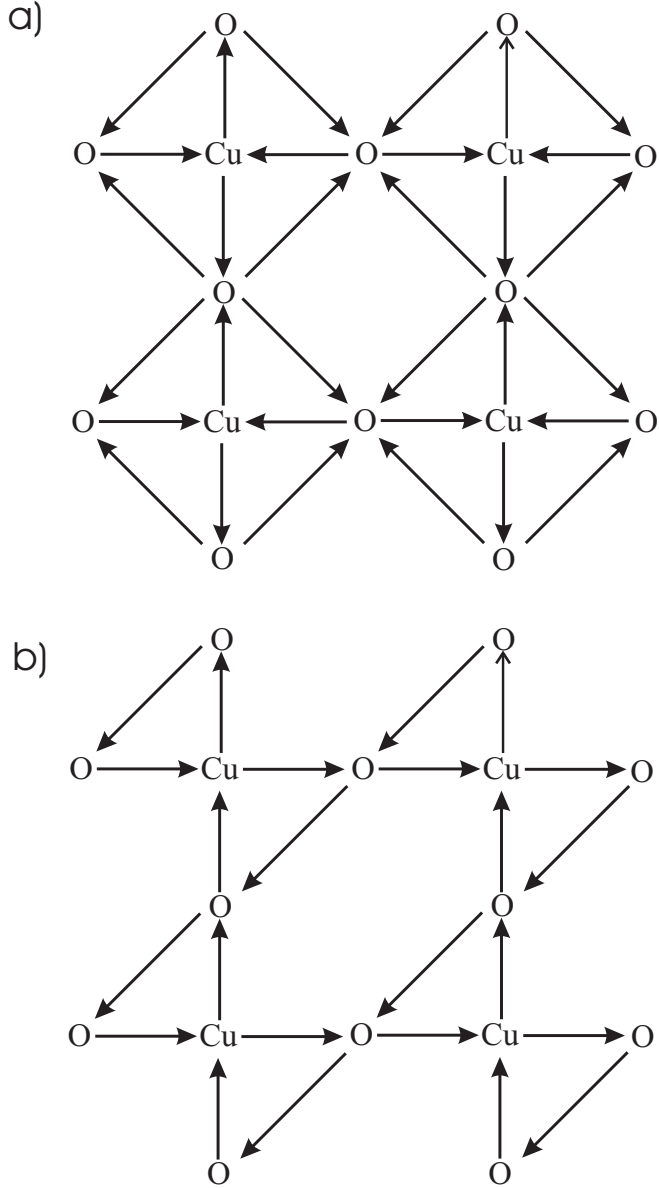}
\caption[]{\protect{Current patterns for the time-reversal breaking states proposed in  Ref. \cite{cmv1,simon}.}}
\label{fig:1}
\end{figure}
       
           The SHG intensity  is $I \propto |S|^2$. As shown below,
           interesting effects arise in experimental geometries in which both $\chi^{e(c)}$
and $\chi^m$ are used and interfere. Then effects linear in the magnetoelectric order parameter
are observable. If light is propogated in the sample along any of the axes, 
$\nabla \times {\bf M} = 0$. Then only the electric-dipole susceptibility contributes
to the SHG and $I \propto |\chi^e|^2$ which is in turn proportional only to
 the square of the order parameter.

          We  fix the axes with respect to the crystal as in Fig. (2).
 For an arbitrary direction of incidence,
        Figure 2 shows the triad form by the propagation vector ${\bf q}$ 
(specified by the polar angle $\theta$ and the azimuthal angle $\phi$), 
and the components of the polarization along the orthogonal directions
$\hat{e}_{xy}$ in the x-y plane  and $\hat{e}_{qz}$ in the q-z plane.
           To maximize the second term in Eq. (\ref{s}) we chose a polar 
incidence angle $\theta=\pi/4$. 

\begin{figure}[hhh]
\centering
\includegraphics[]{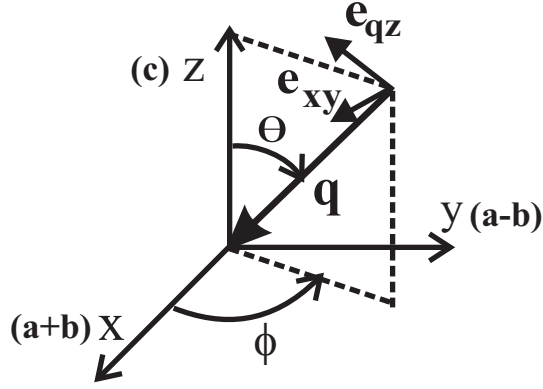}
\vspace{.5 cm}
\caption[]{\protect{Geometry of incident light vector {\bf q}. The axes are fixed 
to the crystal: $\hat{x}$ and $\hat{y}$ are rotated
      $\pi/4$ with respect to the crystalline axis a and b as explained in the text.}}
\label{fig:2}
\end{figure}

          For light linearly polarized along the direction ${\bf E}=E \hat{e}_{qz} $,
          \begin{equation}
          {\bf S}=E^2 \omega \left( \begin{array}{c} 
            \omega \sin{\phi} s^e_1\\
            \omega \cos{\phi} s^e_2 -  |{\bf q}| \sin{\phi}\cos{\phi}
              (\chi^m_x+\chi^m_y+\chi^m_z)/(2\sqrt{2}) \\
             \omega \sin{\phi} s^e_3-|{\bf q}|(\cos^2{\phi} \chi^m_y+\sin^2{\phi} \chi^m_x)/{2} 
                \end{array} \right)
          \end{equation}
          where,
          \begin{eqnarray}
           s^e_1 & = & (\cos{\phi} (\chi_{1}^{e(c)} -\chi_{2}^{e(c)}) -\cos^2{\phi} \chi_{xxy}^{e(c)}
                       -\sin{\phi}^2\chi_{yyy}^{e(c)}-\chi_{zzy}^{e(c)} )/(2 \sqrt{2})\\ \nonumber
           s^e_2 & = & -(\chi_{zzy}^{e(c)}+\cos^2{\phi} \chi_{xxy}^{e(c)}
                       +\sin^2{\phi}(\chi_{yyy}^{e(c)}-\chi_{1}^{e(c)}))/2 \\ \nonumber
           s^e_3 & = & (\cos{\phi} (\chi_{1}^{e(c)} +\chi_{2}^{e(c)}) -\cos^2{\phi} \chi_{xxy}^{e(c)}
                       -\sin{\phi}^2\chi_{yyy}^{e(c)}-\chi_{zzy}^{e(c)} )/(2 \sqrt{2})       
          \end{eqnarray}

           The component are given in the basis $({\bf \hat{q}}, \hat{e}_{xy} ,\hat{e}_{qz})$. 
           \begin{eqnarray}
           I & \propto & E^4 \omega^2 \left[ \omega^2 (\sin^2{\phi}(|s^e_1|^2+|s^e_3|^2 )+\cos^2{\phi}|s^e_2|^2) \right. \\ \nonumber
             &+ & |{\bf q}|^2( |\sin{\phi}\cos{\phi}(\chi^m_x+\chi^m_y+\chi^m_z)|^2/8+
                          |\cos^2{\phi} \chi^m_y+\sin^2{\phi} \chi^m_x|^2) \\ \nonumber
             &- &\left. \omega |{\bf q}| \sin{\phi} (\cos^2{\phi} \Re{ s^e_2 (\chi^m_x+\chi^m_y+\chi^m_z)^*} 
           / {2\sqrt{2}} + \Re{ s^e_3 (\cos^2{\phi} \chi^m_y+\sin^2{\phi} \chi^m_x)^*)}/2 \right]
           \label{I1}
           \end{eqnarray} 
               The $\chi$'s defined below are complex and $\chi^*$ denotes complex conjugation.
               The last term in Eq. (10) is linear in both $\chi^m$ and $\chi^e$ and is
          therefore proportional to the order parameter.  
   
       For an incident light linearly polarized,
          ${\bf E}=E \hat{e}_{xy} $,
          \begin{equation}
          {\bf S}=E^2 \omega \left( \begin{array}{c} 
            \omega \sin{\phi}  s^e_1 \\
            \omega \cos{\phi} s^e_2 + |{\bf q}| \sin{\phi}\cos{\phi} \chi^m_z / \sqrt{2}\\
             \omega \sin{\phi} s^e_1 \\
                \end{array} \right)
          \end{equation}
           where 
         \begin{eqnarray}
           s^e_1 & = & (\sin{\phi}\cos{\phi}\chi_{1}^{e(c)}
                   -\sin^2{\phi} \chi_{yyy}^{e(c)}-\cos^2{\phi}\chi_{xxy}^{e(c)})/\sqrt{2} \\ \nonumber
           s^e_2 & = &-( \sin^2{\phi}(\chi_1^{e(c)}+\chi_{xxy}^{e(c)})+\cos^2{\phi}\chi_{yyy}^{e(c)}          
          \end{eqnarray}

          It is interesting  to consider  circularly polarized light with right 
          $E_{+}=-1/\sqrt{2}(E_{xy}+i E_{zq})$ and left $E_{-}=1/\sqrt{2}(E_{xy}-i E_{zq})$  polarizations
          and evaluate the intensity $I \propto |{\bf S}|^2$ for each case. This gives
          \begin{eqnarray}
          I  & \propto & |E_\pm|^4 \omega^2 \left(\omega^2 |\psi(\phi)^e|^2  + |{\bf q}|^2 |\psi(\phi)^m|^2
          \pm \cos(\phi) Re( i \omega |{\bf q}| \psi_{int}(\phi)^e (\psi_{int}(\phi)^m)^* ) \right)  
          \end{eqnarray}

          Again the last term is the interesting term as it is linear in both $\chi^m$ and $\chi^e$ and is
          therefore proportional to the order parameter. Its sign is changed on reversing 
          the polarization of the light
          or by reversing time (i.e. by reversing the domains or reversing the direction of propagation for
        a fixed domain i.e. changing $\phi$ to $\pi+\phi$).
          For $\cos{\phi}=\pm 1$, that is when ${\bf q}$ belongs to the broken symmetry plane $\hat{y}$,
       the difference between both circular polarization is maximized. For this case,
          \begin{eqnarray}
           |\psi(0)^e|^2 & = & |\chi_2^{e(c)} |^2+|\chi_1^{e(c)} |^2+|\chi_{yyy}^{e(c)} |^2+
          \frac{1}{2} |\chi_{xxy}^{e(c)} +\chi_{zzy}^{e(c)} |^2 \\ \nonumber
           2|\psi(0)^m|^2 & = &   |\chi_y^m|^2+|\chi_x^m+\chi_z^m|^2    \\ \nonumber
           \psi_{int}(\pi/2)^e (\chi_{int}(\pi/2)^m)^* & = & (\chi_{xxy}^{e(c)} 
           +\chi_{zzy}^{e(c)}-2\chi_{yyy}^{e(c)} )(\chi_z^m+\chi_x^m)^* 
             + (\chi_1^{e(c)} +\chi_3^{e(c)} ) (\chi_2^m)^*
          \end{eqnarray}

         In summary we have determined the SHG intensities, as a function of polarization and  direccion 
of the incident light, for different group symmetries corresponding to the
         normal and  time-reversal breaking symmetry states.    
           In all the states there is a contribution to the SHG coming from the 
magnetic-dipole terms but the electric-dipole contribution appears
         only when the inversion and time-reversal are broken with their 
 product preserved, as in the 
        $\Theta_{II}$ phase.  
         The {\it interference} of these two terms produce linear terms in both $\chi^m$ and $\chi^e$ 
          and therefore proportional to the order parameter.   
          This term change sign reversing the domains $\Theta_{II}$ to $-\Theta_{II}$ or 
          changing the direction of incident light by the operation ${\bf q}
\rightarrow \sigma_x\sigma_y{\bf q}$ 
(which has the effect of changing$\phi$ to $\phi+\pi$).
          The {\it interference} term also changes sign on changing from left
 to right circularly polarized light. 
               
          The presence of many different $\Theta_{II}$ domains within the 
spot size of the incident beam will blur this effect. ARPES experiments {\cite{kaminski}
indicate domains to be of O(100 microns).

The proposed experiment are a rather thorough test of the symmetries of
 the predicted phase because they rely on broken R and $\sigma_y$ (or $\sigma_x)$) 
but their product preserved below the pseudogap temperature in the cuprates. Breaking reflection
symmetry alone would not produce the predicted effects in Second-harmonic generation, while
the ARPES experiment\cite{kaminski} is in principle compatible \cite{simon} with breaking some reflection symmetries alone.
Some other experiments using resonant x-ray absorption and diffraction have also been proposed (\cite{dimatteo}).
 
{\it {Acknowledgments}}: We wish to thank Professor Harry Tom for discussion 
of the feasibility of experiments and
Professor J. Orenstein who asked the questions which led to this investigation.

           \end{document}